# Effect of Meissner screening and trapped magnetic flux on magnetization dynamics in thick Nb/Ni$_{80}$Fe$_{20}$/Nb trilayers

Kun-Rok Jeon,[1,2] Chiara Ciccarelli,[2] Hidekazu Kurebayashi,[3] Lesley F. Cohen,[4] Xavier Montiel,[5] Matthias Eschrig,[5] Thomas Wagner,[1,6] Sachio Komori,[1] Anand Srivastava,[1] Jason W. A. Robinson,[1] and Mark G. Blamire[1]

[1]*Department of Materials Science and Metallurgy, University of Cambridge, 27 Charles Babbage Road, Cambridge CB3 0FS, United Kingdom*

[2]*Cavendish Laboratory, University of Cambridge, Cambridge CB3 0HE, United Kingdom*

[3]*London Centre for Nanotechnology and Department of Electronic and Electrical Engineering at University of College London, London WC1H 01H, United Kingdom*

[4]*The Blackett Laboratory, Imperial College London, SW7 2AZ, United Kingdom*

[5]*Department of Physics, Royal Holloway, University of London, Egham Hill, Egham, Surrey TW20 0EX, United Kingdom*

[6]*Hitachi Cambridge Laboratory, Cambridge CB3 0HE, United Kingdom*

We investigate the influence of Meissner screening and trapped magnetic flux on magnetization dynamics for a Ni$_{80}$Fe$_{20}$ film sandwiched between two thick Nb layers (100 nm) using broadband (5-20 GHz) ferromagnetic resonance (FMR) spectroscopy. Below the superconducting transition $T_c$ of Nb, significant zero-frequency line broadening (5-6 mT) and DC resonance field shift (50 mT) to a low field are *both* observed if the Nb thickness is comparable to the London penetration depth of Nb films ($\geq$ 100 nm). We attribute the observed peculiar behaviors to the increased incoherent precession near the Ni$_{80}$Fe$_{20}$/Nb interfaces and the effectively **focused magnetic flux in the middle $Ni_{80}Fe_{20}$ caused by strong Meissner screening and (defect-)trapped flux of the thick adjacent Nb layers. This explanation is supported by static magnetic properties of the samples and comparison with FMR data on thick $Nb/Ni_{80}Fe_{20}$ bilayers. Great care should therefore be taken in the analysis of FMR response in ferromagnetic Josephson structures with thick superconductors, a fundamental property for high-frequency device applications of spin-polarized supercurrents.**

## I. INTRODUCTION

In the past two decades, a ferromagnetic Josephson junction (FJJ) comprising two superconductors (SCs) separated by a ferromagnet (FM) has been of interest and developed extensively because of its unconventional physical properties [1-3] and potential applications in cryogenic computing technologies [4-10]. Very recent experimental work has demonstrated that nanotextured FJJs integrated with standard single flux quantum neural systems form a new class of neuromorphic technologies that have spiking energies of less than $10^{-18}$ J, operation frequencies up to 100 GHz, and nanoscale plasticity [11]. In particular for an emergent field of superconducting spintronics [8-10], it has been recently established that the presence of a spatially varying magnetization $M(x)$ at SC/FM interfaces can generate spin-polarized triplet supercurrents via spin mixing and spin rotation processes into the FM [12-14]. Interestingly, almost a decade ago, theoretical studies [15,16] suggested a time-varying magnetization $M(t)$ of the FM as a reciprocal equivalent to $M(x)$ for the generation of spin-polarized triplet supercurrents in a diffusive metallic FJJ [15] and also in a FM/NM/SC structure (NM: normal metal) [9].

However, subsequent ferromagnetic resonance (FMR) studies on Nb/Ni$_{80}$Fe$_{20}$ bilayers [17,18] and Nb/Ni$_{80}$Fe$_{20}$/Nb trilayers [19] have shown that spin angular momentum transfer in such structures is predominantly mediated by quasiparticles (QPs) for the superconducting state and thus largely suppressed at a lower temperature $T$ by the development of singlet superconductivity and the freeze-out of available QP states [17,20,21]. This is likely because the magnitude of $M(t)$ inhomogeneity or non-collinearity, parameterized by the magnetization precession angle $\theta_M$, is too small (a few degrees at 10-20 GHz) [17-19] to yield the measurable effect of $M(t)$-induced triplet supercurrents [15,16]. Another recent experiment, on the other hand, has reported that for Ni$_{80}$Fe$_{20}$ films sandwiched between rather thick Nb layers (100 nm) [22], the DC resonance field shifts remarkably to a low field below the superconducting transition temperature $T_c$, interpreted as possible evidence for field-like spin-transfer torque (STT) induced by spin-triplet supercurrents.

According to the STT theory for metallic magnetic heterostructures [23-25], the anti-damping STT is expected to be much larger (an order of magnitude) than the field-like STT due to the rapid dephasing of transverse spins in the FM [25]. It is thus of fundamental importance to test whether the effect of the anti-damping torque (relevant to the Gilbert damping change $\Delta[\alpha]$) is consistent with that of the field-like torque (associated with the resonance field shift $\Delta[\mu_0 H_{res}]$) in the superconducting state. Furthermore, knowledge about how magnetization dynamics of the FM changes by Meissner screening and magnetic flux pinning [26,27], especially in contact with thick SC layers, is highly desirable for the successful implementation of FMR functionality in FJJ-based superconducting spintronics [8-10]. Note that the triplet proximity channel which is required to carry spin angular momentum depends on the strength of the underlying singlet superconductivity and thus thicker superconducting electrodes are favorable for the generation of higher density superconducting spin currents in FJJs [8-10, 12-14].

Here, we focus on thick Nb/Ni$_{80}$Fe$_{20}$/Nb trilayers where the Nb thickness $t_{Nb}$ is comparable to the London penetration depth $\lambda_L$ of Nb films ($\geq$ 100 nm) [28] and so there exists a non-negligible effect of Meissner screening on the local (DC/RF) magnetic field experienced by the middle Ni$_{80}$Fe$_{20}$ layer. Through broad-band (5-20 GHz) FMR measurements on such trilayers, we identify that the anomalous zero-frequency line broadening $\mu_0\Delta H_0$ and the significant $\Delta[\mu_0 H_{res}]$ to a low field *both* appear below $T_c$. Importantly, the effect of $\Delta[\mu_0 H_{res}]$ is found to be 1-2 orders of magnitude larger than that of $\Delta[\alpha]$, which is incompatible with the STT theory [23-25]. We explain these peculiar behaviors in terms of locally perturbed magnetization precession of the middle Ni$_{80}$Fe$_{20}$ layer under spatially inhomogeneous magnetic fields caused by the strong Meissner effect and the magnetic flux pinning [26,27] of the thick adjacent Nb layers. Static magnetic properties of the samples and comparison with FMR data on thick Nb/Ni$_{80}$Fe$_{20}$ bilayers consistently support our explanation.

## II. EXPERIMENTAL DETAILS

Polycrystalline Nb/Ni$_{80}$Fe$_{20}$/Nb trilayers and Nb/Ni$_{80}$Fe$_{20}$ bilayers are deposited on thermally oxidized Si substrates with lateral dimensions of 5 mm × 5 mm using DC magnetron sputtering in an ultra-high vacuum chamber. The Nb (Ni$_{80}$Fe$_{20}$) thickness $t_{Nb}$ ($t_{Py}$) of 100 (15) nm is chosen to allow comparison with the recent FMR study on similar sample structures [22]. Details of the sample growth and $T_c$ characterization are described elsewhere [19].

We measure the FMR response of the sample attached on a broadband coplanar waveguide with either DC field or RF pulse modulation [19]; to obtain each FMR spectrum, the absorbed microwave (MW) power by the sample is measured while sweeping the external static magnetic field $\mu_0 H$ at the fixed MW frequency $f$ of 5 to 20 GHz. Note that for all FMR measurements, the MW power is set to 10 dBm where the actual MW power absorbed in the sample is a few mW that has no effect on $T_c$ of the Nb layer [19]. At the beginning of each measurement, we apply a large in-plane $\mu_0 H$ (0.5 T) to fully magnetize the $Ni_{80}Fe_{20}$ layer, after which the field is reduced to the range of FMR. Once the $f$-dependent FMR measurements (from high- to low-$f$) finish, the field is returned to zero to cool the system down further for a lower $T$ measurement. We employ a vector field cryostat from Cryogenic Ltd that can apply a 1.2 T magnetic field in any direction over a $T$ range of 2−300 K. Some FMR measurements are conducted on the same samples using a different Helium flow cryostat from Oxford Instruments to test for reproducibility.

Magnetization properties of the same samples used for FMR measurements are characterized using a Quantum Design Magnetic Property Measurement System at $T$ varying between 2 and 300 K. For all FMR and magnetization measurements, $\mu_0 H$ is applied parallel to the film plane; a careful alignment of the film plane with respect to $\mu_0 H$ is made to minimize any unintentional out-of-plane component of $\mu_0 H$.

## III. RESULTS AND DISCUSSION

### A. Temperature dependence of ferromagnetic resonance at different frequencies

Let us first consider the $T$ evolution of FMR spectra for the Nb(100 nm)/$Ni_{80}Fe_{20}$(15 nm)/Nb(100 nm) trilayer. Figure 1(a) shows typical FMR data obtained at the two different $f$ of 10 and 20 GHz, from 80 K down to 2 K. Note that all the FMR data presented are well fitted with the field derivative of symmetric and antisymmetric Lorentzian functions [29]. This enables us to accurately determine the FMR linewidth $\mu_0\Delta H$ (linked to the Gilbert damping $\alpha$) and the resonance field $\mu_0 H_{res}$ (associated with the saturation magnetization $\mu_0 M_s$). Overall $T$ dependences of $\mu_0\Delta H$ and $\mu_0 H_{res}$ for various $f$ are summarized in Figs. 1(b) and 1(c), respectively. In the normal state ($T > T_c$), $\mu_0\Delta H$ and $\mu_0 H_{res}$ are both almost independent of $T$. However, on entering the superconducting state ($T < T_c$), $\mu_0\Delta H$ broadens largely down to 4 K followed a slight fall at a lower $T$ and $\mu_0 H_{res}$ shifts significantly to a low field; these effects are more pronounced for a lower $f$. This superconducting state FMR response is quite different from observed in the relatively thin Nb/Ni$_{80}$Fe$_{20}$/Nb samples ($t_{Nb} \leq 60$ nm $<< \lambda_L$) where $\Delta[\mu_0 H_{res}]$ is less than 2% (at $f = 20$ GHz) and $\mu_0\Delta H$ narrows monotonically below $T_c$ [19], implying that the $t_{Nb}$-dependent superconductivity itself is responsible for the difference between them.

**B. Significant zero-frequency line broadening and resonance field shift below $T_c$**

For a quantitative analysis, we extract the Gilbert(-type) damping constant $\alpha$ from the linear scaling of $\mu_0\Delta H$ with $f$ at a fixed $T$ [Fig. 2(a)]; $\mu_0\Delta H(f) = \mu_0\Delta H_0 + \frac{4\pi\alpha f}{\sqrt{3}\gamma}$ [30]. Here $\mu_0\Delta H_0$ is the zero-frequency line broadening due to long-range magnetic inhomogeneities in the FM [31] and $\gamma$ is the gyromagnetic ratio ($1.84 \times 10^{11}$ T$^{-1}$ s$^{-1}$) [32]. In the formula, we exclude other extrinsic broadening effects such as two-magnon scattering [33,34] and Mosaicity broadening [34,35] as these have non-linear $f$ and weak $T$ dependences. The extracted $\alpha$ [Fig. 2(b)] progressively decreases deep into the superconducting state ($T < T_c$), which can be explained by the suppressed outflow of spin currents from the precessing Ni$_{80}$Fe$_{20}$ due to the development of singlet superconductivity in the adjacent Nb layers. This agrees with theoretical studies for proximity-coupled metallic FM/SC systems [18] and also with previous experiments [17, 19-21].

In contrast, an anomalous $\mu_0 \Delta H_0$ [inset of Fig. 2(b)] appears for the thick Nb/Ni$_{80}$Fe$_{20}$/Nb sample ($t_{Nb}$ = 100 nm) below $T_c$ and it reaches 5-6 mT at 2 K, approximately an order of magnitude stronger than that in the relatively thin Nb/Ni$_{80}$Fe$_{20}$/Nb samples ($t_{Nb} \leq 60$ nm $\ll \lambda_L$) [19]. This implies that when $t_{Nb}$ is comparable to $\lambda_L$ [28], the coupled superconducting Nb layers perturb locally magnetization precession of the Ni$_{80}$Fe$_{20}$ layer and cause the incoherent precession near the Ni$_{80}$Fe$_{20}$/Nb interfaces.

The influence of superconductivity on $\mu_0 H_{res}(f)$ can be described using the modified Kittel formula [36]: $f = \frac{\gamma}{2\pi}\sqrt{[\mu_0(H_{res} + H_{shift}^{SC} + M_{eff}) \cdot \mu_0(H_{res} + H_{shift}^{SC})]}$, where $\mu_0 M_{eff}$ is the effective saturation magnetization and $\mu_0 H_{shift}^{SC}$ is the correction term that describes the superconductivity-induced resonance field shift. In Fig. 2(c), we fit the $\mu_0 H_{res}(f)$ data obtained at different (constant) $T$ using the Kittel formulas with and without the presence of $\mu_0 H_{shift}^{SC}$ for comparison. When $\mu_0 H_{shift}^{SC} \neq 0$, all the data are well fitted [Fig. 2(c)] and the estimated values of $\mu_0 M_{eff}$ are in the range of 835-850 mT [Fig 2(d)], which are similar to those obtained from static magnetometry measurements (Sec. C). By contrast, when $\mu_0 H_{shift}^{SC} = 0$, fitting the data gets worse at a lower $T$ particularly for a lower $f$ [Fig. 2(c)] and gives the anomalously increased $\mu_0 M_{eff}$ below $T_c$ [Fig. 2(d)]. This points to that there exist an internal source of DC magnetic flux/field to the middle Ni$_{80}$Fe$_{20}$, accompanied by the onset of superconductivity in the Nb layers.

Notably, the extracted $\mu_0 H_{shift}^{SC}$ of 30-50 mT [inset of Fig. 2(d)] is found to be 1-2 orders of magnitude larger than the FMR damping decrease of 0.0005-0.0045 for the superconducting state [Fig. 2(b)], corresponding to the $\mu_0 \Delta H$ suppression of 0.4-3.5 mT at 20 GHz (in the dimension of effective field). Since this result is inconsistent with the STT theory [23-25] described above, it is natural to consider an alternative explanation, more relevant to the superconductivity-modified magnetization dynamics. The most common feature of SC films is the presence of Meissner screening and magnetic flux pinning [26, 27].

### C. Static magnetic properties below $T_c$

To support our explanation of the FMR result, we perform static magnetometry measurements on the same samples (used for FMR measurements) across $T_c$. Figure 3(a) first shows the magnetization versus $T$ plots for the Nb(100 nm)/Ni$_{80}$Fe$_{20}$(15 nm)/Nb(100 nm) trilayer and the Nb(100 nm)/Ni$_{80}$Fe$_{20}$(15 nm) bilayer. FMR data of the bilayer and their comparison with the trilayer will be presented below (Sec. D). Above $T_c$ of the Nb, the total magnetization $M_{total}$ of the sample is given solely by the ferromagnetic Ni$_{80}$Fe$_{20}$ layer, which is expected to increase weakly with decreasing $T$ (far below $T_{Curie}$) as $1 - B \cdot T^{\frac{3}{2}}$ according to Bloch's law [37]. Here $B$ is the Bloch constant or spin-wave parameter. A fair fit (black solid line) to the data [inset of Fig. 3(a)] is obtained with a reasonable $B$ of 1.25 (1.26) × 10$^{-5}$ K$^{-3/2}$ for the trilayer (bilayer), which is very close to the estimated value ($B$ = 1.23 × 10$^{-5}$ K$^{-3/2}$) for bulk Ni$_{80}$Fe$_{20}$ [38]. On the other hand, below $T_c$, the Nb layers with type-II SC magnetization can contribute to $M_{total}$ (of the sample) in addition to the ferromagnetic Ni$_{80}$Fe$_{20}$ - indicative of this is an abrupt change in $M_{total}$ (under the field cooling) when $T_c$ is crossed.

The $M_{total}$ versus in-plane $\mu_0 H$ curves across $T_c$ are presented in Figs. 3(b) and 3(c) for the trilayer and the bilayer, respectively. Assuming that the superconducting state $M_{total}(\mu_0 H)$ is a superposition of ferromagnetic $Ni_{80}Fe_{20}$ and Nb (type-II SC) magnetizations, one can separate the Nb magnetization $M_{Nb}(\mu_0 H)$ by taking the difference between the $M_{total}(\mu_0 H)$ curves above and below $T_c$. We then get the characteristic type-II behavior in initial curves [insets of Figs. 3(b) and 3(c)]. The linear diamagnetic response to $\mu_0 H$ (Meissner screening) is visible for a small field range ($\leq 0.2$ Tesla). After reaching an extremum at the lower critical field $\mu_0 H_{c1}$, the absolute magnetization drops as magnetic flux starts to penetrate the Nb until reaching the upper critical field $\mu_0 H_{c2}$ (a few Tesla for Nb films) [19,39].

It is notable that magnetic flux pinning at defects in the SC can be inferred from the hysteresis behaviors, which emerge when $\mu_0 H > \mu_0 H_{c1}$. The hysteresis area and the remaining magnetization at zero external field, quantifying the amount of flux pinning, are both expected to be much larger for thicker SCs where more defect sites and stronger Meissner screening co-exist [26,27]. The consistent behaviors seen in the $M_{Nb}(\mu_0 H)$ curves [insets of Figs. 3(b) and 3(c)] clarify that the non-negligible (defect-)trapped magnetic flux is present in the thick Nb samples.

In fact, the anomalous FMR response observed in the thick Nb/$Ni_{80}Fe_{20}$/Nb trilayer below $T_c$ (Sec. B) can be explained if we consider that the trapped magnetic flux at defects randomly distributed in the neighboring Nb layers serves as the internal source of additional magnetic field to the middle $Ni_{80}Fe_{20}$ under the external DC resonance field.

### D. Comparison with bilayers

To further support our explanation, let us now discuss the FMR results (Fig. 4) taken from the Nb(100 nm)/$Ni_{80}Fe_{20}$(15 nm) bilayer where overall flux pinning (of the sample) is weaker compared to the trilayer (Sec. C). Figures 4(a)-4(c) show that for the bilayer, the change of FMR spectra, i.e. $\mu_0\Delta H$ and $\mu_0H_{res}$, as function of $T$ below $T_c$ is indeed weaker than for the trilayer. Note that what $\mu_0\Delta H$ tends to increase at a lower $T$ [Fig. 4(b)] means is the occurrence of superconductivity-induced line broadening, as discussed below.

For better understanding and quantitative comparison, we plot the $T$ dependences of $\alpha$, $\mu_0H_0$, $\mu_0M_{eff}$, and $\mu_0H_{shift}^{SC}$ values in Figs. 4(d) and 4(e), extracted from the $f$-dependent FMR data (see Appendix for details). It is worth noting that the quantitative change of $\alpha$ reduction (or spin-pumping damping) across $T_c$ [Fig. 4(d)] is approximately *twice weaker* relative to the trilayer [Fig. 2(b)], which is in accordance with spin pumping through a single Nb/Ni$_{80}$Fe$_{20}$ interface [17,18] and thereby a single spin-blocking effect of the Nb layer [17-21].

Perhaps, the most noteworthy aspect of the bilayer data is that even though the resonance field shift is very small [< |0.5| mT, inset of Fig. 4(e)], as in the previous experiment [16], there still exists the anomalous increase of $\mu_0H_0$ at a lower $T$ [+2 mT at 3.3 K, inset of Fig. 4(d)] that is approximately 3 times smaller than the trilayer [inset of Fig. 2(b)]; but large enough to compensate the FMR linewidth suppression [−2.5 mT at 3.3 K at 20 GHz, Fig. 4(d)] resulting from the aforementioned spin-blocking effect [17-21]. It in turn makes the $T$ dependence of total linewidth nontrivial [Fig. 4(b)], highlighting that broad-band FMR measurements are of utmost importance for proper interpretation of the experimental results.

The bilayer result suggests that the FMR linewidth change is more sensitive than the resonance field shift to the local flux pinning and so the $f$-dependent linewidth analysis may be useful to isolate somehow the genuine spin-triplet proximity effect [8-10] from other extrinsic phenomena [40, 41] being driven in FM/SC interfaces, a key ingredient for developing superconducting spintronics.

## IV. CONCLUSIONS

How Meissner screening and (defect-)trapped magnetic flux affect magnetization dynamics in thick Nb/Ni$_{80}$Fe$_{20}$/Nb trilayers is investigated by using broadband FMR spectroscopy. We find that when $t_{Nb}$ is comparable to $\lambda_L$ of Nb films, anomalous $\mu_0 \Delta H_0$ and significant $\Delta[\mu_0 H_{res}]$ to a low field *both* appear below $T_c$. Notably, the effect of $\Delta[\mu_0 H_{res}]$ is found to be much greater than that of $\Delta[\alpha]$ in the superconducting state, which is incompatible with the STT theory. We consider the superconductivity-modified magnetization dynamics as an alternative explanation for the FMR data, which is convincingly supported by static magnetic properties of the samples and comparison with FMR data on thick Nb/Ni$_{80}$Fe$_{20}$ bilayers. Our results suggest that careful consideration should be made when analyzing FMR data in FJJs with thick SCs. Proper selection of SC properties provides a pathway to dynamically access the spin-polarized supercurrents in SC/FM proximity-coupled systems [8-10] for their potentials in high-frequency device applications.

## ACKNOWLEDGMENTS

This work was supported by EPSRC Programme Grant EP/N017242/1.

## APPENDIX A: ANALYSIS OF FREQUENCY DEPENDENCE OF FERROMAGNETIC RESONANCE SPECTRA FOR THE BILAER

Using the same approach as for the trilayer (Sec. B), we extract the *T* dependences of $\alpha$, $\mu_0 H_0$, $\mu_0 M_{eff}$, and $\mu_0 H_{shift}^{SC}$ values [presented in Figs. 4(d) and 4(e)] from the *f*- dependent FMR data acquired on the thick Nb(100 nm)/Ni$_{80}$Fe$_{20}$(15 nm) bilayer (Fig. 5). Note that there exists the visible increase of $\mu_0 H_0$ (zero-frequency intercept) at a lower $T$ [Fig. 5(a)] even though the resonance field shift is tiny [Fig. 5(b)].

## APPENDIX B: DISCUSSION OF MEISSNER SCREENING EFFET ON MAGNETIC DOMAIN STRUCTURE

It was previously reported that for disk-patterned Al(150 nm)/Ni(50 nm) samples of submicron size in zero external field [42], Meissner screening of ferromagnetic domain's stray fields by the adjacent Al layer can cause a spatial re-distribution of magnetic domains and its $T$ dependence is connected with the screening capability of the Al. Even if this mechanism would, in principle, increase magnetization inhomogeneity below $T_c$, it is unclear that this could be used to explain our results, where continuous samples are used in FMR study under application of a large external field (30-400 mT). In this vein, it is of interest to systematically investigate how the superconducting state FMR response is altered by changing the type of FM (having a different strength of stray fields) in FM/SC bilayer and SC/FM/SC trilayer structures.

magnetization changes and nonlocal effects in mesoscopic ferromagnet-superconductor structures, Phys. Rev. B **65**, 220513(R) (2002).

## FIGURE CAPTIONS

FIG. 1. (a) Typical FMR spectra for the thick Nb(100 nm)/Ni$_{80}$Fe$_{20}$(15 nm)/Nb(100 nm) trilayer obtained at the two different frequencies *f* of 10 and 20 GHz, from 80 K down to 2 K. The yellow (blue) background represents the normal (superconducting) state of Nb. Temperature *T* dependence of the FMR linewidth $\mu_0\Delta H$ (b) and the resonance magnetic field $\mu_0 H_{res}$ (c) for the Nb/ Ni$_{80}$Fe$_2$/Nb trilayer. The inset shows the normalized resistance $R/R_N$ versus *T* plot for the trilyaer. Note that at *f* = 5 GHz in (b), FMR signals become unmeasurable when *T* < 8 K as the amplitude of the resonance field shift to a low field goes beyond the typical resonance field (~33 mT) in the normal state. It is also noteworthy that in (b), the peak point is determined by two competing effects of 1) Meissner screening and flux pinning and 2) spin-blocking behavior (see main text for details).

FIG. 2. (a) FMR linewidth $\mu_0\Delta H$ as a function of microwave frequency for the Nb(100 nm)/Ni$_{80}$Fe$_{20}$(15 nm)/Nb(100 nm) trilayer at various temperatures *T*. The solid lines are linear fits to deduce the Gilbert damping constant *α* and the zero-frequency line broadening $\mu_0\Delta H_0$. (b) Deduced values of *α* and $\mu_0\Delta H_0$ (inset) as a function of *T*. (c) Microwave frequency versus resonance field $\mu_0 H_{res}$. The solid lines are fits to extract the effective saturation magnetization of the Ni$_{80}$Fe$_{20}$ layer using the modified Kittel formulas with (left) and without (right) the correction term $\mu_0 H_{shift}^{SC}$. (d) Extracted $\mu_0 M_{eff}$ values (of the Ni$_{80}$Fe$_{20}$) versus *T* with and without the presence of $\mu_0 H_{shift}^{SC}(T)$ (inset).

FIG. 3. (a) Total magnetization $M_{total}$ versus temperature $T$ plots for the Nb(100 nm)/Ni$_{80}$Fe$_{20}$(15 nm)/Nb(100 nm) trilayer and the Nb(100 nm)/Ni$_{80}$Fe$_{20}$(15 nm) bilayer. $M_{total}(T)$ is attained while decreasing $T$ at the fixed/applied magnetic field $\mu_0 H$ of 8 mT, which is far less than the lower critical field $\mu_0 H_{c1}$ of Nb layers. The $M_{total}$ versus (in-plane) magnetic field $\mu_0 H$ curves, taken above and below the superconducting transition $T_c$ for the Nb/Ni$_{80}$Fe$_{20}$/Nb trilayer (b) and the Ni$_{80}$Fe$_{20}$/Nb bilayer (c). The diamagnetic background signal from the quartz sample holder is subtracted. Each inset shows the isolated Nb magnetization $M_{Nb}(\mu_0 H)$ by taking the difference between the $M_{total}(\mu_0 H)$ curves above and below $T_c$. The arrow in the inset is a guide to the eyes for the initial curve.

FIG. 4. (a) Representative FMR spectra for the thick Nb(100 nm)/Ni$_{80}$Fe$_{20}$(15 nm) bilayer obtained at the fixed frequency $f$ of 10 GHz, from 80 K down to 2 K. The yellow (blue) background represents the normal (superconducting) state of Nb. Temperature $T$ dependence of the FMR linewidth $\mu_0 \Delta H$ (b) and the resonance magnetic field $\mu_0 H_{res}$ (c) for the Nb/ Ni$_{80}$Fe$_2$ bilayer. The inset exhibits the normalized resistance $R/R_N$ versus $T$ plot for bilayer (d) Estimated values of $\alpha$ and $\mu_0 \Delta H_0$ (inset) as a function of $T$. (e) Extracted $\mu_0 M_{eff}$ values (of the Ni$_{80}$Fe$_{20}$) versus $T$ with and without the presence of $\mu_0 H_{shift}^{SC}(T)$ (inset). Relevant details are presented in Appendix A.

FIG. 5. Data equivalent to Figs. 2(a) and 2(c) but for the Nb(100 nm)/Ni$_{80}$Fe$_{20}$(15 nm) bilayer.

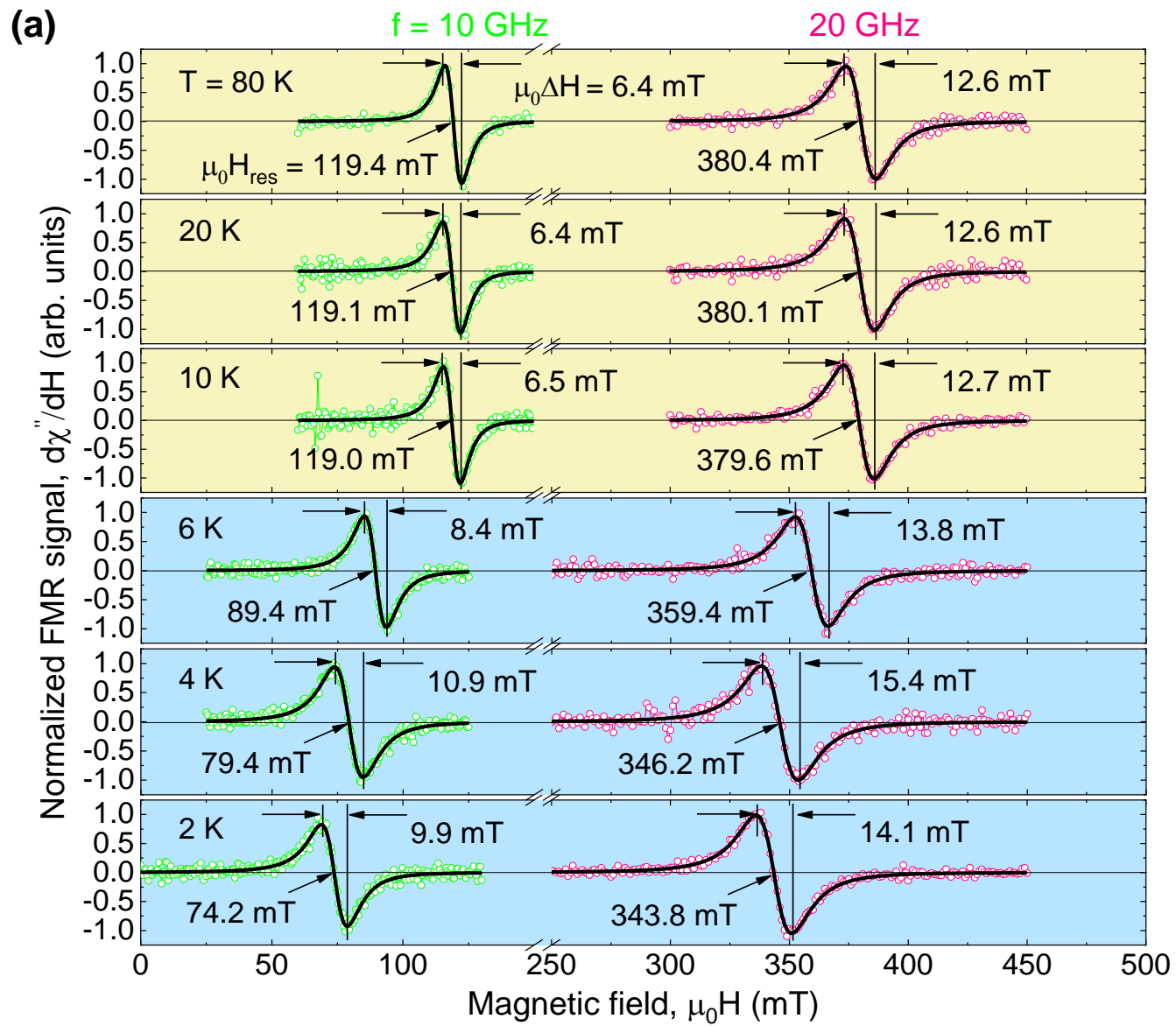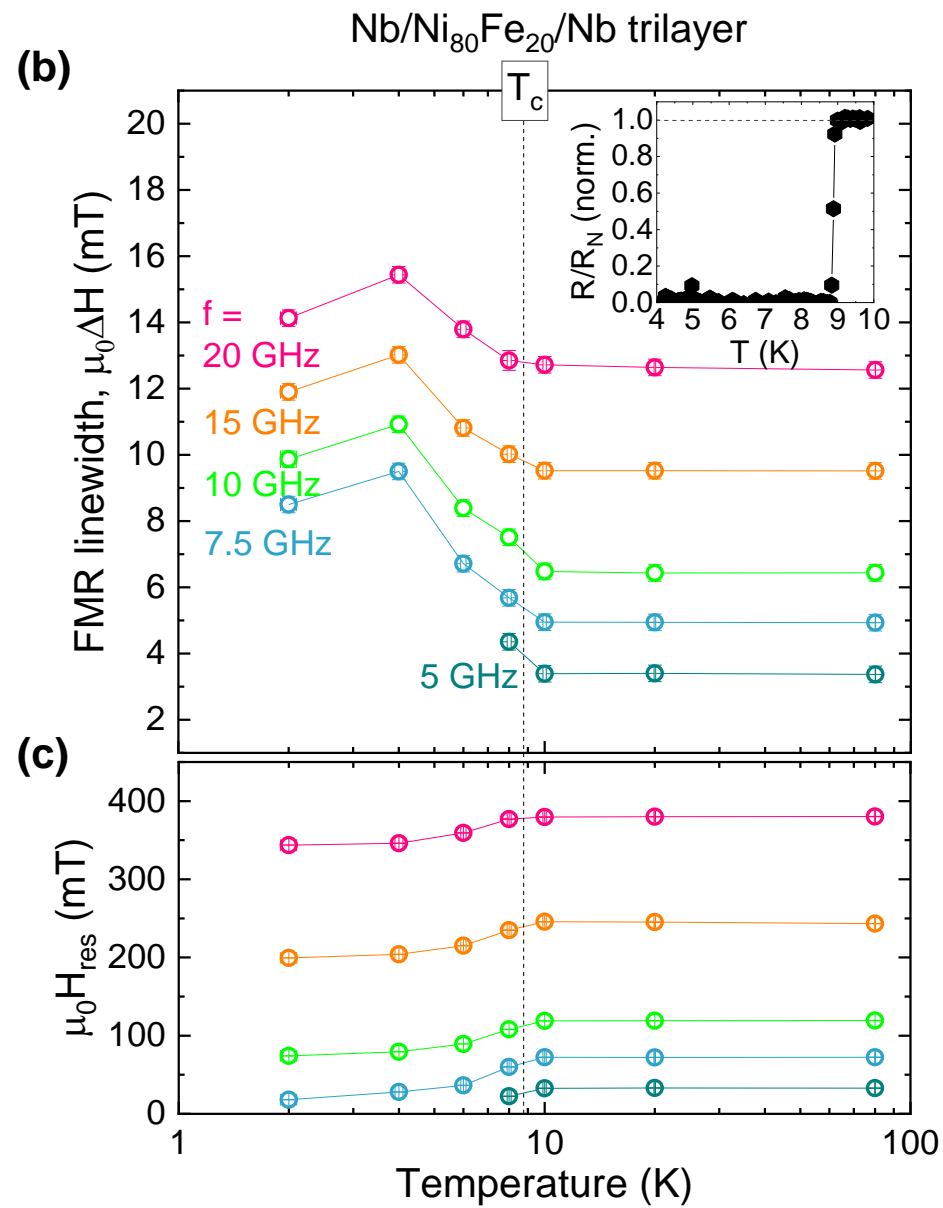

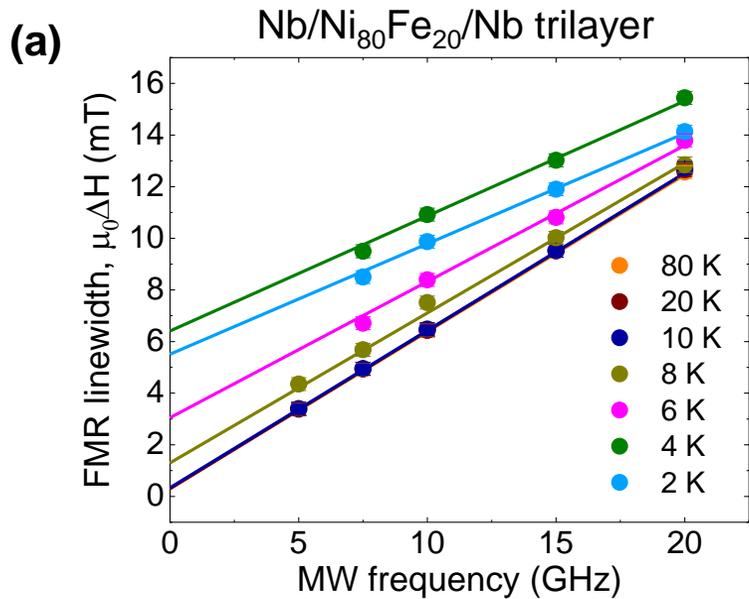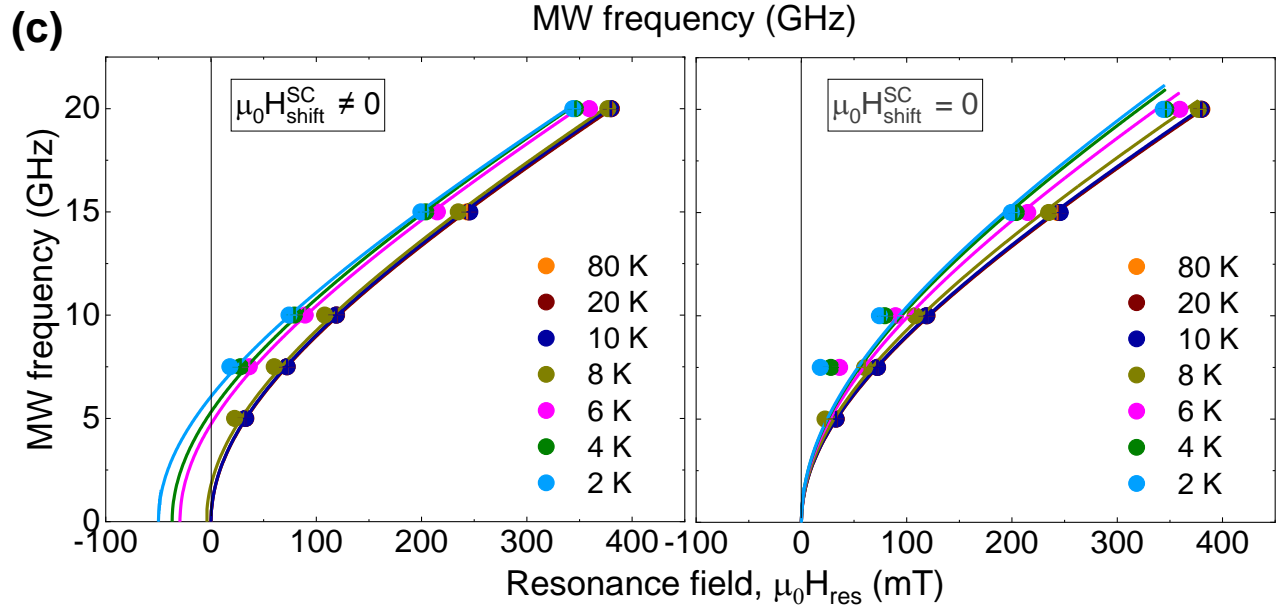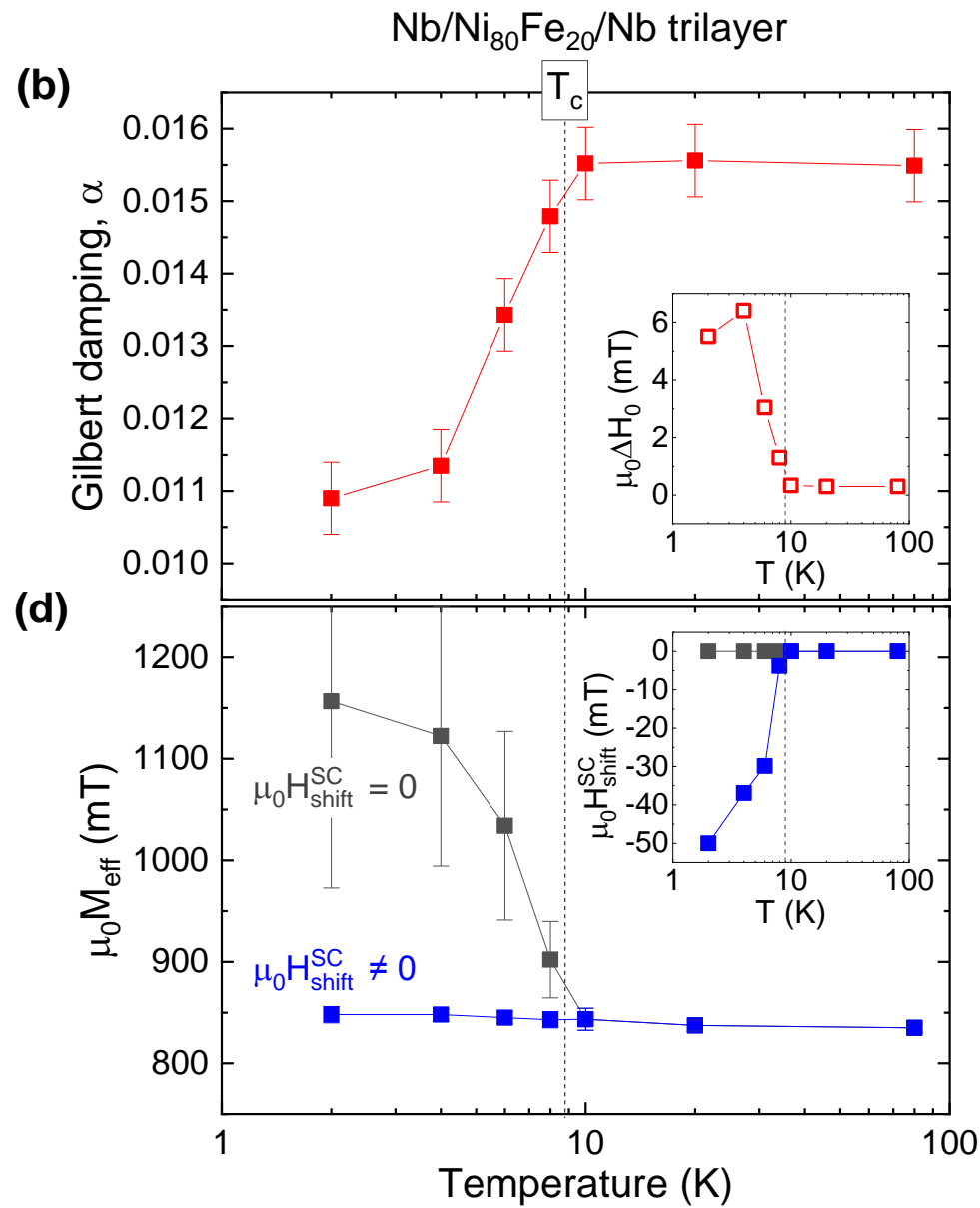

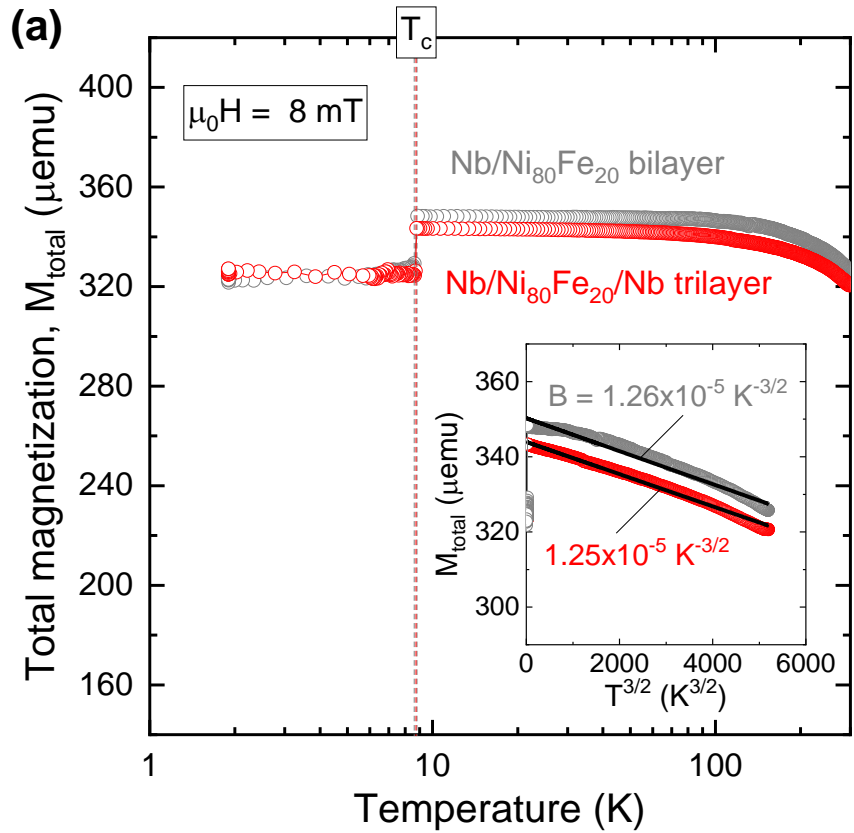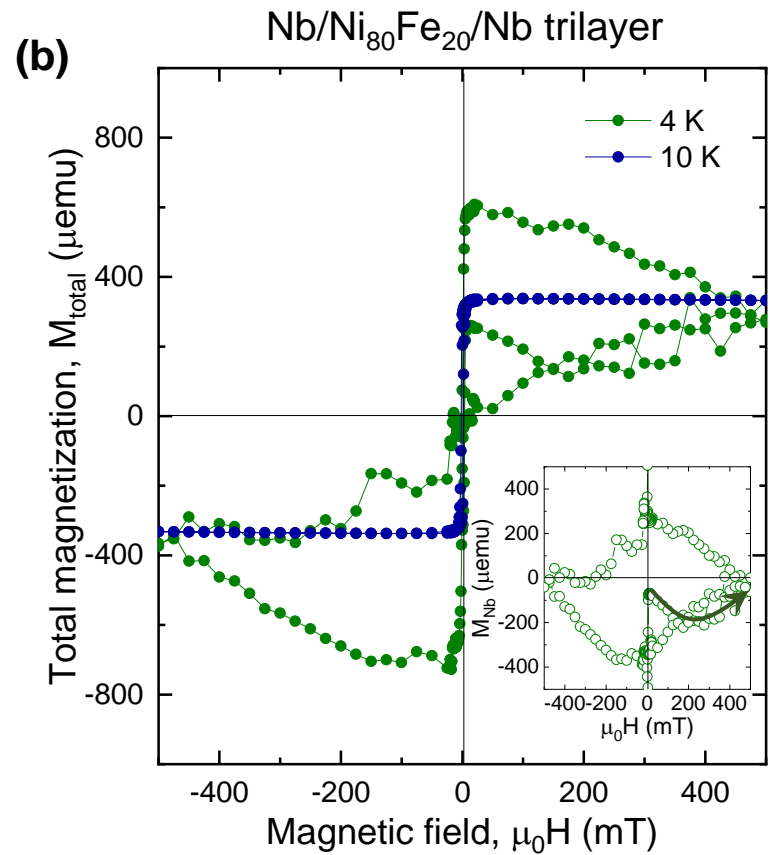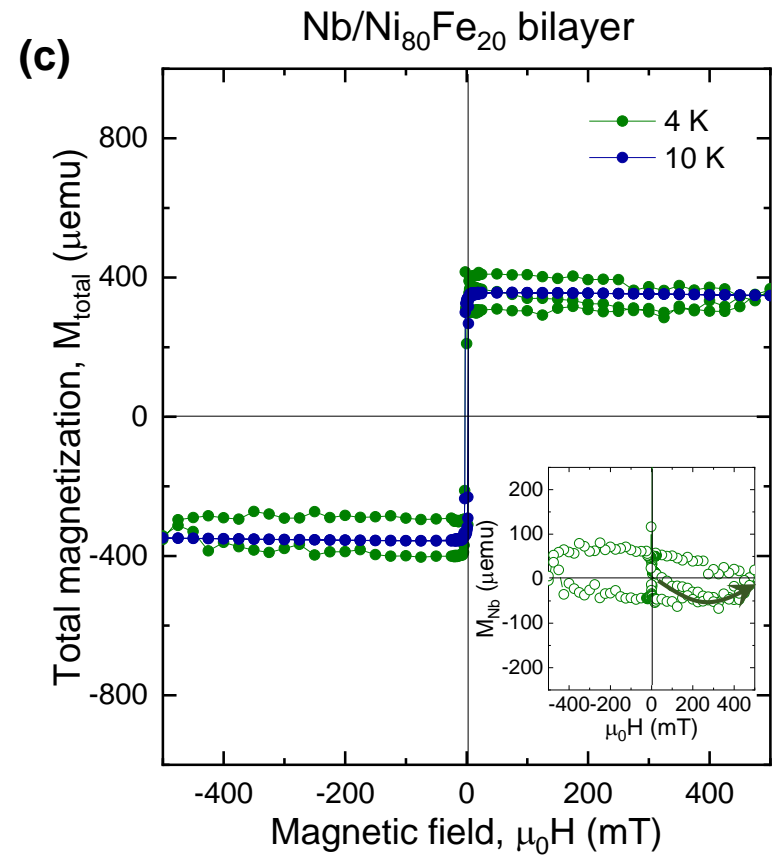

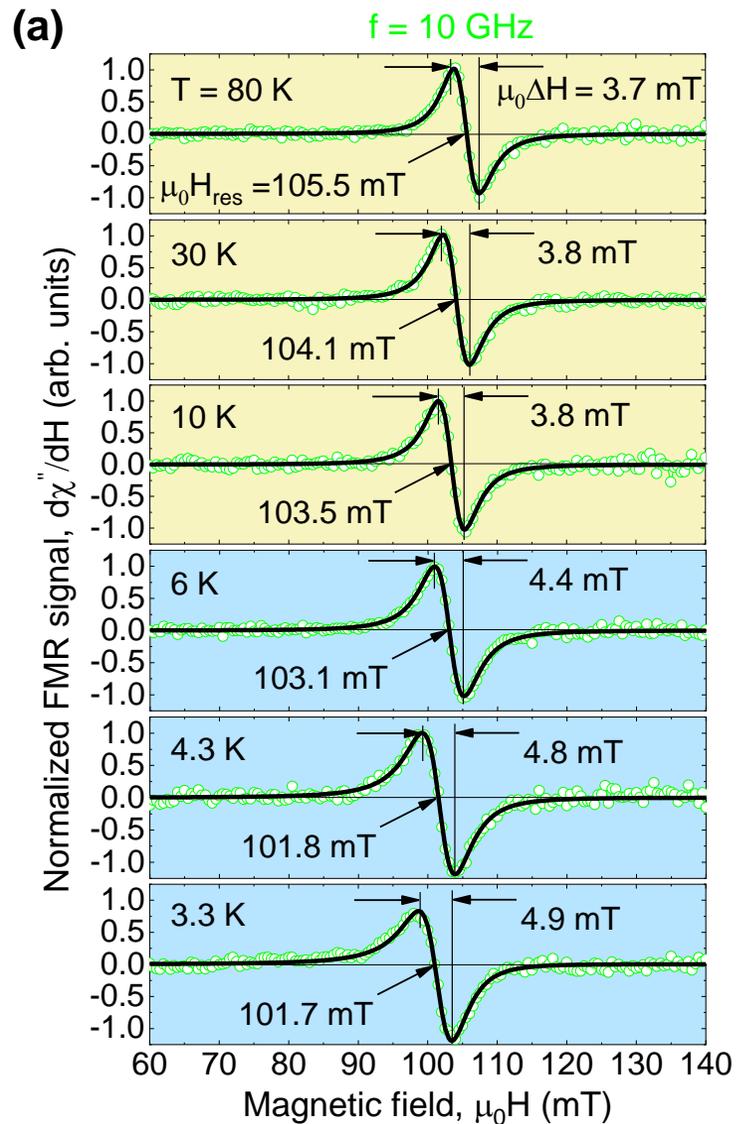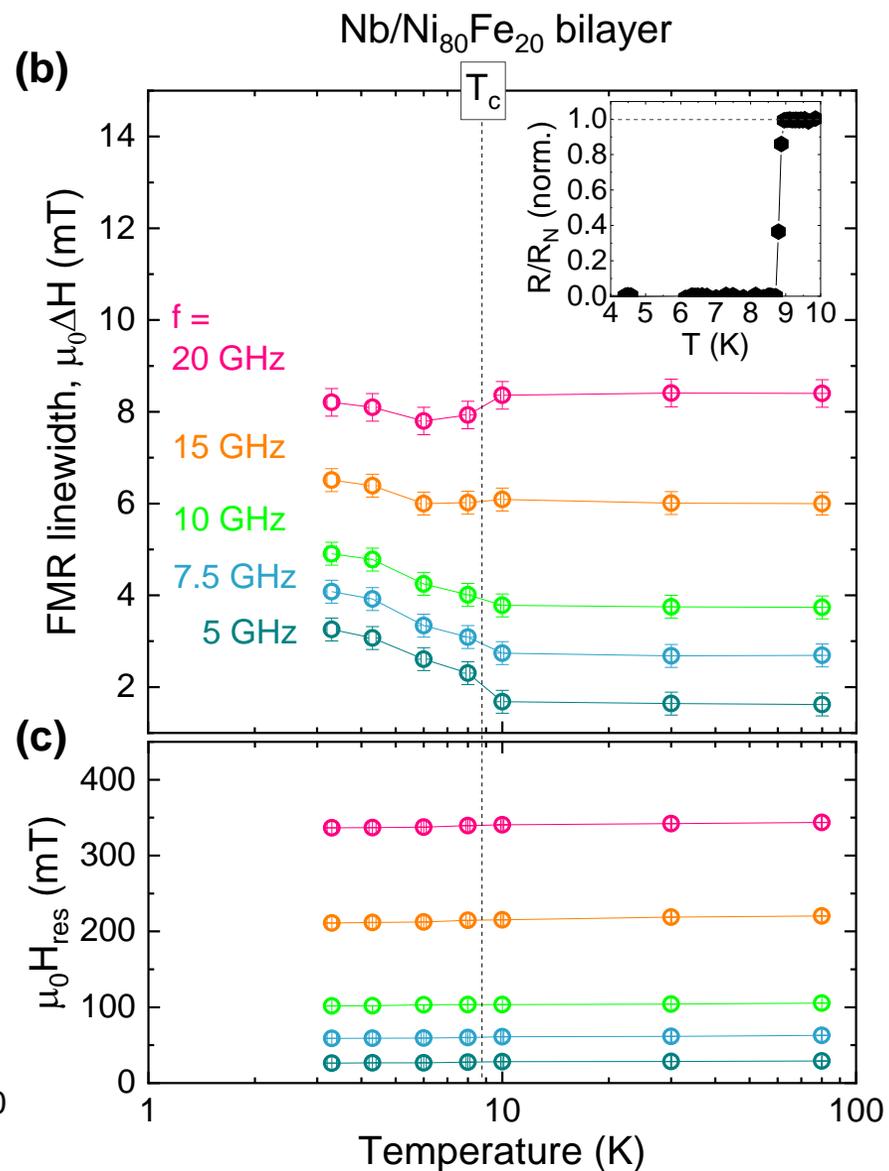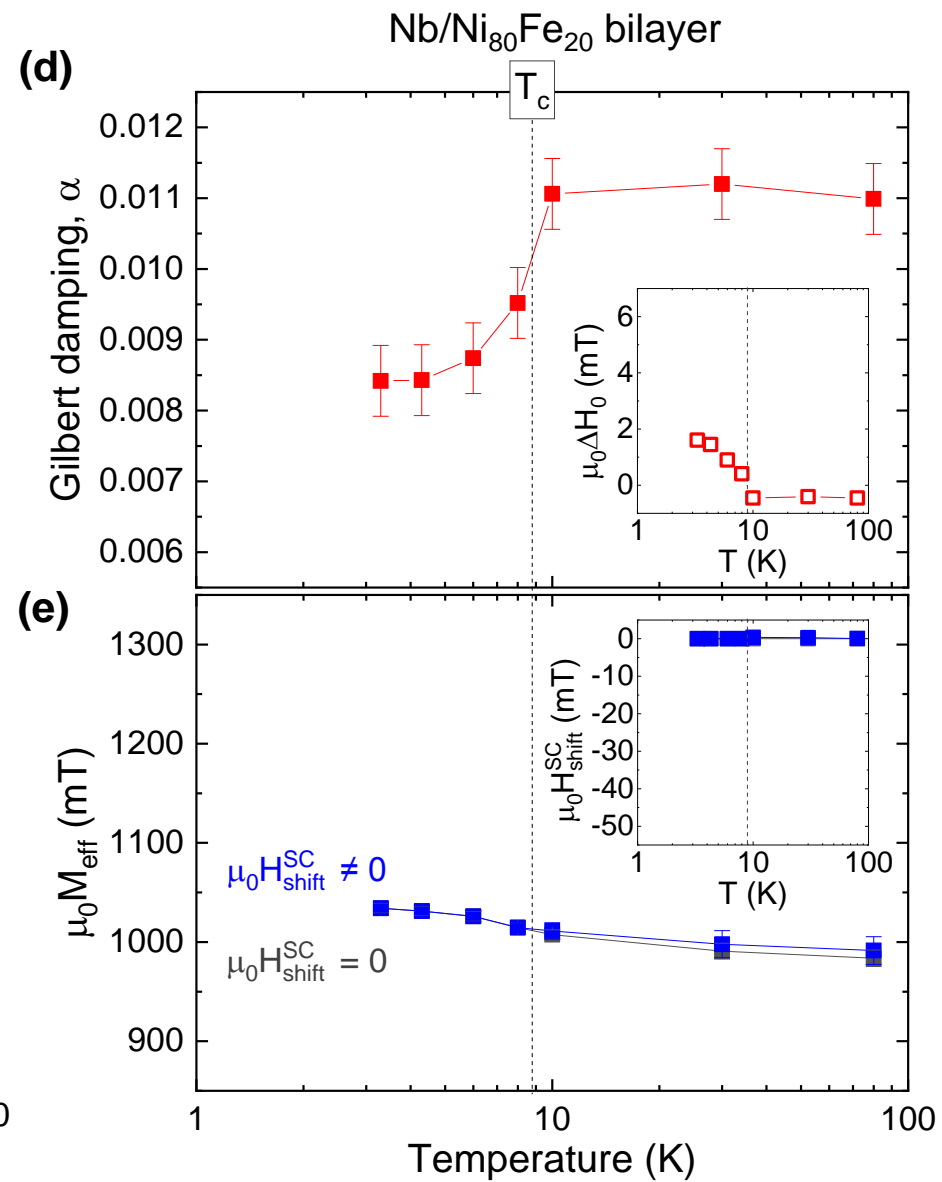

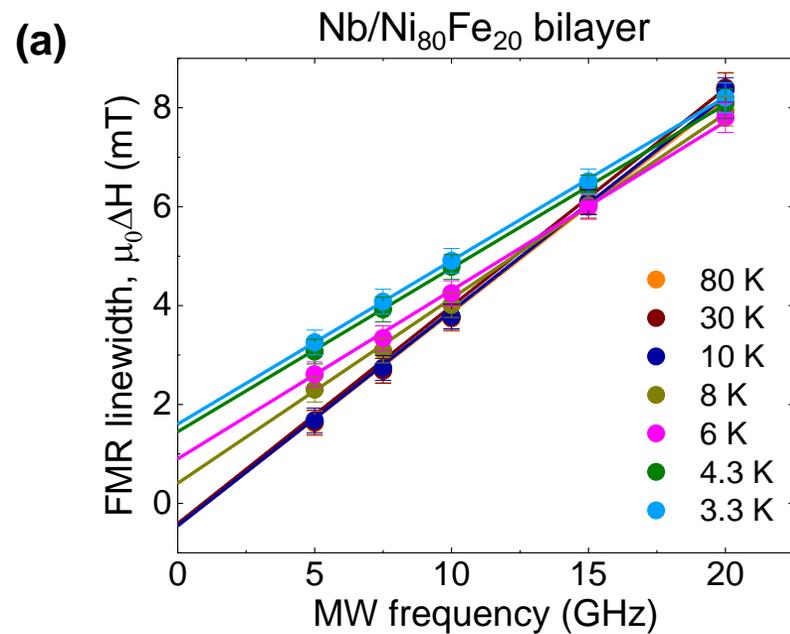
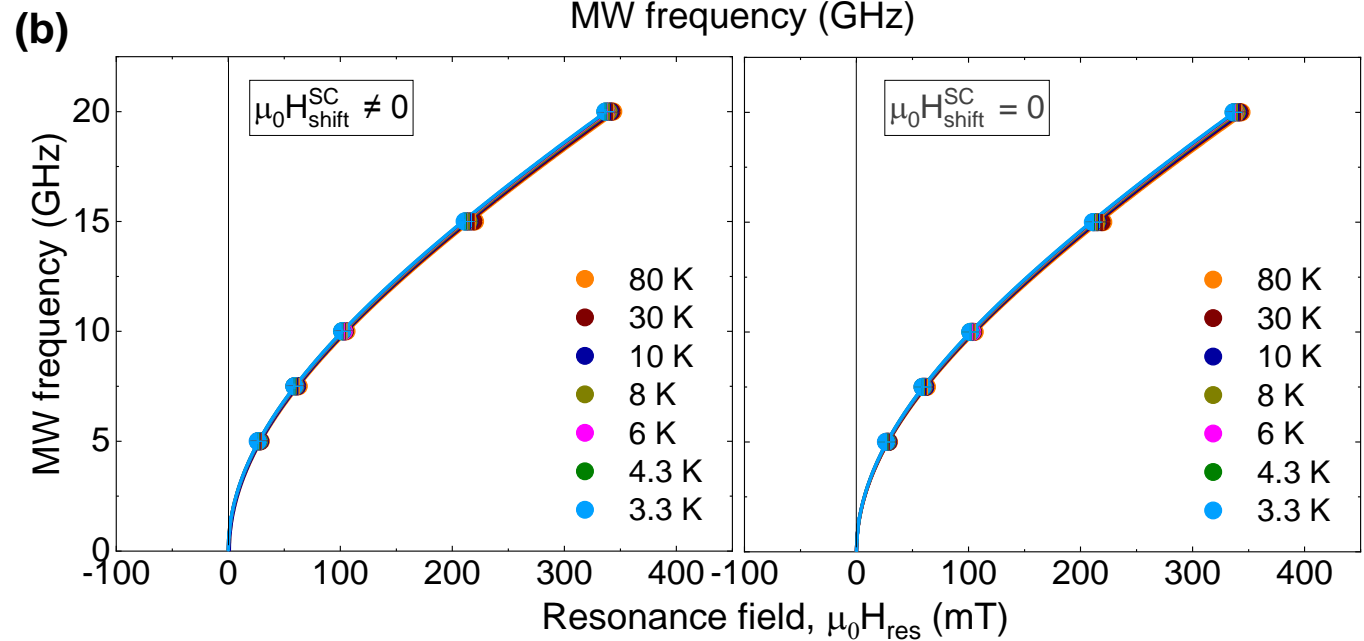